\documentclass[12pt]{article}

\usepackage{graphicx,amssymb}

\newcommand{\tr}{\mbox{Tr} \, }
\newcommand{\rank}{\mbox{rank} \,}
\newcommand{\partialtr}[1]{\mbox{Tr}_{#1} \,}
\newcommand{\ket}[1]{\left | #1 \right \rangle}
\newcommand{\bra}[1]{\left \langle #1 \right |}
\newcommand{\amp}[2]{\left \langle #1 \left | #2 \right. \right \rangle}

\newcommand{\proj}[1]{\ket{#1} \! \bra{#1}}

\newcommand{\cpmap}[2]{{\cal E}^{#1}_{#2}}
\newcommand{\unity}{\mbox{\bf I}}
\newcommand{\hilbert}{{\cal H}}

\newcommand{\bigsum}[1]{{\displaystyle \sum_{#1}}}
\newcommand{\subspace}[1]{{\cal S}_{#1}}
\newcommand{\noinfo}{\not \rightsquigarrow}

\begin{document}

\title{Locality and information transfer in quantum operations}

\author{Benjamin Schumacher$^{(1)}$
        and Michael D. Westmoreland$^{(2)}$}
\maketitle
\begin{center}
{\sl
$^{(1)}$Department of Physics, Kenyon College, Gambier, OH 43022 USA \\
$^{(2)}$Department of Mathematical Sciences, Denison University,
 Granville, OH  43023 USA }
\end{center}

\section*{Abstract}

We investigate the situation in which no information can be
transferred from a quantum system $B$ to a quantum system $A$,
even though both interact with a common system $C$.

\section{Introduction}

The universe can be divided up into subsystems that interact with
one another.  All parts of the universe are connected, directly or
indirectly, by this web of interactions.  Nevertheless, to predict
the future state of a small subsystem $A$, it is not necessary
to specify the past state of the whole universe.  This is what we
mean by ``locality'' of the dynamical evolution of $A$ within
the global system.

Beckman et al. \cite{preskill} have investigated a related notion of
locality in the context of quantum operations.  Suppose we have
a bipartite system $AB$ whose quantum state evolves according
to the map $\cpmap{AB}{}$.  We say that this map is {\em semicausal}
if it cannot be used to transfer information from $B$ to $A$.
That is, if we begin with a joint state $\rho^{AB}$, perform
an operation $\cal B$  on subsystem $B$, and finally
apply the map $\cpmap{AB}{}$ to the joint system then the
final state of $A$ alone is independent of the choice of $\cal B$.
A {\em causal} map is semicausal in both directions.
In \cite{preskill}, these notions are related to other
more constructive properties of the map $\cpmap{AB}{}$.
Roughly speaking, we say that the map is {\em semilocalizable}
if it can be represented as successive interactions with a
common ancilla system $R$:  first $A$ interacts with $R$ and
then $B$ interacts with $R$.  The map is {\em localizable}
if it is semilocalizable in both directions.  Because of the
order of these interactions, it can be seen that a semilocalizable
map is also semicausal.  Beckman et al. give an example of
a map that is fully causal but not localizable.
In \cite{werner} it is further shown
that all semicausal maps are semilocalizable.

However, the framework of \cite{preskill} and \cite{werner} does
not seem sufficiently general to capture the notion of locality.
From the outset, it is assumed that the joint system $AB$ is
effectively isolated.  (While it is true that the map $\cpmap{AB}{}$
may include interaction with an external environment,
a knowledge only of the past state of $AB$ itself is
sufficient to predict the future state of $AB$.)
Furthermore, if $\cpmap{AB}{}$ is semicausal, then $A$ itself
is also effectively isolated---that is, there exists a map
$\cpmap{A}{}$ that yields future $A$ states given only past
$A$ states as input.
In other words, the future state of $AB$ is determined
by the past state of $AB$, and no influence can propagate
from $B$ to $A$ during the time interval.
But there are many situations in which these things are
not true, but for which we would say that the dynamics is local.

\begin{figure}[htbp]
\begin{center}
\includegraphics[height=1.5in]{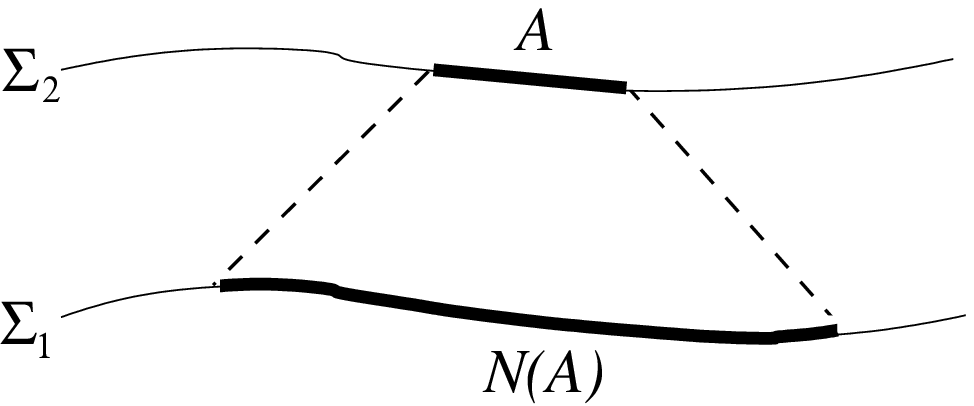}
\end{center}
\caption{In spacetime, the state of a field in the region $A$ on the
    later hypersurface $\Sigma_2$ only depends on the state of the field
    in the region $N(A)$ on the earlier hypersurface $\Sigma_1$.
    \label{cauchy-fig}}
\end{figure}

For example, suppose we are considering the dynamics of a classical
relativistic field $\phi$ in spacetime.  ``Moments of time''
are spacelike hypersurfaces in our spacetime.  The state of
$\phi$ in a region $A$ of one hypersurface is completely
determined by the state of $\phi$ in a somewhat larger region
$N(A)$ of an earlier hypersurface.\cite{cauchy}
The dynamics of this field
is local, inasmuch as we can ignore the rest of the universe
outside of $N(A)$ when predicting the future field configuration
on $A$.  Yet we cannot find two nonempty spatial regions $A$ and $B$
so that (1) the future joint field state on $AB$ is
determined only by the past field state on $AB$, and (2) no
influence can propagate from $B$ to $A$ during the time interval.
See Figure~\ref{cauchy-fig}.
The definition of semicausality cannot capture the notion of locality
for the evolution of this kind of system.

\begin{figure}[htbp]
\begin{center}
\includegraphics[height=1.5in]{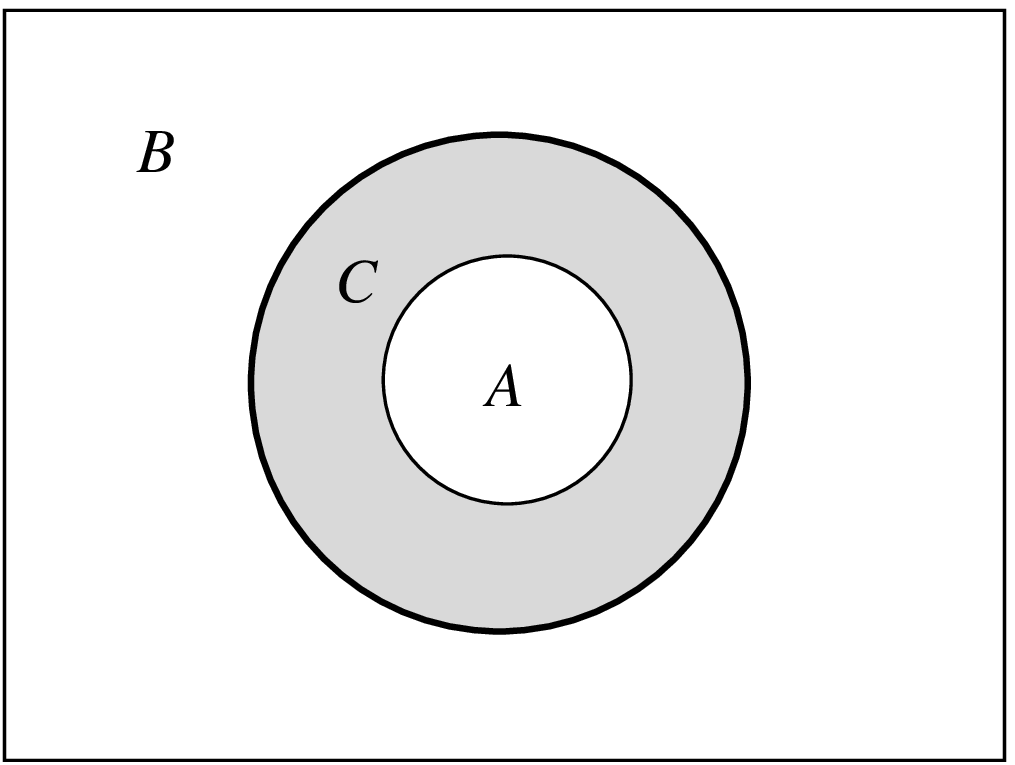}
\end{center}
\caption{Three ``concentric'' systems.
    \label{3system1-fig}}
\end{figure}

We need an idea of locality based on a division of the universe into
{\em three} subsystems.  See Figure~\ref{3system1-fig}, in which these
subsystems are represented by concentric planar regions.
Subsystem $A$ is surrounded by subsystem $C$,
which includes the rest of the dynamical ``neighborhood''
of $A$.  We call $C$ the {\em context} of $A$.
To predict the final state of $A$, we only need to
know the initial state of the composite system $AC$.
Beyond $A$ and its context is subsystem $B$, which contains
the rest of our universe, and whose state is irrelevant
to the final state of $A$.

Because the initial state of $B$ does not affect the final
state of $A$, no {\em information transfer} is possible
from $B$ to $A$ under the dynamical evolution.  We write
this condition as $B \noinfo A$.

In this paper we aim, first, to make precise the dynamical notion
of locality in quantum mechanics and to clarify its relation to
information transfer.  Second, we will use these ideas to explore
what sort of local dynamics is possible if the global quantum evolution
is unitary.

\section{Heuristics for quantum dynamical maps}

We begin by reviewing some results about the dynamics of closed
and open quantum systems.  In a closed system, the evolution of
the quantum state is described by a unitary operator $U$.  An
initial pure state vector $\ket{\psi}$ evolves to a final pure
state vector $\ket{\psi'}$ according to
\begin{equation}
    \ket{\psi} \longrightarrow \ket{\psi'} = U \ket{\psi} .
\end{equation}
If instead we describe the initial state by a density operator
$\sigma$, the final state is described by
\begin{equation}
    \sigma \longrightarrow \rho = U \sigma U^{\dagger}.
\end{equation}

An open quantum system interacts with its surroundings, and this
interaction can lead to noise and decoherence in its time evolution.
A more general description of this evolution would be a map $\cpmap{}{}$
from initial to final density operators---that is,
\begin{equation}
    \sigma \longrightarrow \rho = \cpmap{}{} \left ( \sigma \right ) .
\end{equation}
What properties must the map $\cpmap{}{}$ possess?  It clearly must
be trace-preserving, since $\tr \sigma = \tr \rho = 1$.  (We will
assume without further comment that all of our maps are trace-preserving.)
Also, $\cpmap{}{}$ must be a positive map, always
taking a positive operator $\sigma$ to a positive operator $\rho$.
Furthermore, it must be {\em completely positive}
(CP), which means that when we extend the map to the map
$\unity \otimes \cpmap{}{}$ on a larger system, it remains positive.
Physically, this means that we can append to our quantum system a
second ``ancilla'' system that has trivial dynamics (described by
the identity map $\unity$), and the overall evolution
of the composite system still takes positive density operators
to positive density operators.


Every CP map $\cpmap{}{}$ has a {\em unitary representation}.  That is,
we can introduce an external ``environment'' system $E$ that is
initially in a standard state $\ket{0}$ and find a unitary operator
$U$ on the composite system such that
\begin{equation}
    \cpmap{}{} \left ( \sigma \right )
        = \partialtr{E} U \, \left ( \sigma \otimes \proj{0} \right )
        \, U^{\dagger}
\end{equation}
for all $\sigma$.
This not only gives a convenient representation
for any CP map, it also makes a crucial physical point about
when such maps are appropriate descriptions.  The
evolution can be described by a CP map only when
the quantum system interacts with an external
system with which it is not initially correlated.
In more general situations where initial correlations
may exist, we cannot treat the external system
as an ``environment'' and derive a local CP map
for the system of interest.

Any CP map $\cpmap{}{}$ also has an {\em operator-sum representation},
which means that there are operators $A_{\mu}$ such that
\begin{equation}
    \cpmap{}{} \left ( \sigma \right ) = \sum_{\mu}
                A_{\mu} \sigma {A_{\mu}}^{\dagger}
\end{equation}
for all $\sigma$.  The operators $A_{\mu}$ satisfy
$\bigsum{\mu} {A_{\mu}}^{\dagger} A_{\mu}  =  1$.  A given
CP map has many different operator-sum representations.

When there is any chance of confusion, we indicate the particular
system to which a state, operator or map applies by a superscript.
Thus, $\ket{\psi^{C}}$ is a pure state vector for $C$,
$X^{AB}$ is an operator for the composite system $AB$,
and $\cpmap{Q}{}$ is a map on $Q$ states.  We will also need
to consider maps {\em between} two distinct systems---in other
words, maps that take states of a system $X$ as input and yield
states of a system $Y$ as output.  We will indicate this using
both superscripts and subscripts, like so:
\begin{equation}
    \rho^{Y} = \cpmap{Y}{X} \left ( \sigma^{X} \right ) .
\end{equation}
(If the CP map is written with no subscript, the input and output
spaces are the same.)  The partial trace operation is a simple
example of this type of map.

To specify a CP map, we would in general need to say how it
acts on many different input states.  However, there is a way
to specify the map by describing the action of its extension
on a single input state.  Let $\cpmap{Q}{}$ be a CP map on
$Q$ states, and let us append an ancilla system $R$ whose
Hilbert space is at least as large as $\hilbert^{Q}$.  The
composite system evolves according to $\unity^{R} \otimes
\cpmap{Q}{}$.  Let $\ket{\Psi}$ be a maximally entangled
state of $RQ$.  Then specifying the output state
\begin{equation}
    \rho^{RQ} =
    \unity^{R} \otimes \cpmap{Q}{} \left ( \proj{\Psi} \right )
\end{equation}
completely specifies the CP map $\cpmap{Q}{}$.  This is a
handy characterization.  If we can show that two CP maps lead
to the same output from a given maximally entangled input,
then we can conclude that the two maps are the same.

We end this section with an observation about the states of
composite systems.  We call a pute state $\ket{\Psi^{RQ}}$ a
{\em purification} of the state $\rho^{Q}$ if
\begin{equation}
    \rho^{Q} = \partialtr{R} \proj{\Psi^{RQ}} .
\end{equation}
A given density operator $\rho^{Q}$ will admit many possible
purifications by $R$, provided $\dim \hilbert^{R}$ is at least
as large as the rank of $\rho^{Q}$.  If $\ket{\Psi^{RQ}_{1}}$
and $\ket{\Psi^{RQ}_{2}}$ are two purifications of the same
state $\rho^{Q}$, then there exists a unitary operator $V^{R}$
on $\hilbert^{R}$ such that
\begin{equation}
    \ket{\Psi^{RQ}_{2}} = \left ( V^{R} \otimes 1^{Q} \right )
        \ket{\Psi^{RQ}_{1}} .
\end{equation}
In other words, any purification of a given state of $Q$
can be turned into any other by the application of a unitary
transformation that only affects the purifying system $R$.

\section{Locality}

How can we express the condition $B \noinfo A$ more precisely?
Let us imagine that $A$, $B$ and $C$ are quantum systems, and that
we have the task of predicting the future state (or the outcomes
of future measurements) of $A$.  The global evolution of the
composite system $ABC$ is described by a CP map $\cpmap{ABC}{}$.

First of all, we can say that $B \noinfo A$ if the future
state of $A$ is a function of the initial quantum state
of the subsystem $AC$ only.  The global initial state is
described by the density operator $\sigma^{ABC}$, but for
making $A$ predictions we only need $\sigma^{AC} = \partialtr{B}
\sigma^{ABC}$.  Call this condition ``Locality (I)'':
\begin{quote}
    {\bf Locality (I).}  There exists a CP map $\cpmap{A}{AC}$ such that,
        \begin{equation}
            \cpmap{A}{AC} \circ \partialtr{B} = \cpmap{A}{ABC}
                    = \partialtr{BC} \circ \cpmap{ABC}{} .  \label{localityI-eq}
        \end{equation}
        That is, for all initial $ABC$-states $\sigma^{ABC}$,
        \begin{equation}
            \cpmap{A}{AC} \left ( \sigma^{AC} \right ) =
            \partialtr{BC} \cpmap{ABC}{} \left ( \sigma^{ABC} \right ) .
        \end{equation}
        To find the final state of subsystem $A$, therefore, it suffices
        to know only the initial state $\sigma^{AC} = \partialtr{B} \sigma^{ABC}$
        of the subsystem $AC$, rather than the global state $\sigma^{ABC}$.
\end{quote}

Alternately, we may focus on the special case when subsystems
$A$, $B$ and $C$ all have definite states to begin with.  In
this case, $B \noinfo A$ means that ignorance of the initial
$B$ state will have no adverse effect on our ability to make
predictions about $A$.  This is ``Locality (II)'':
\begin{quote}
{\bf Locality (II).}  Given pure states $\ket{\gamma}$ of $C$ and
        $\ket{\alpha}$ of $A$, suppose that $\ket{0}$ and $\ket{1}$
        are two pure states (not necessarily
        orthogonal) of $B$.  For $k \in \{ 0,1 \}$, let
        \begin{equation}
            \rho^A_{k} = \partialtr{BC} \cpmap{ABC}{} \left (
                \proj{\alpha} \otimes \proj{k} \otimes \proj{\gamma}
                \right ) .
        \end{equation}
        $B \noinfo A$ means that, for all choices of $\ket{\alpha}$,
        $\ket{\gamma}$ and the $B$-states $\ket{k}$,
        $\rho^{A}_0 = \rho^{A}_1$.
\end{quote}

Finally, $B \noinfo A$ means that no prior
intervention in the $B$ system will affect any
prediction that we make about the future state of $A$ alone.
This is ``Locality (III)'':
\begin{quote}
    {\bf Locality (III)}
        Suppose $ABC$ starts in some arbitrary state $\sigma^{ABC}$,
        and suppose that $\mathcal{F}_{0}$ and $\mathcal{F}_{1}$ are two
        CP maps on $B$ states.  Given $k \in {0,1}$, define
        \begin{equation}
            \rho^{A}_k = \partialtr{BC} \cpmap{ABC}{} \left (
            \mathcal{F}_{k} \otimes \unity^{AC}
            \left ( \sigma^{ABC} \right ) \right ) .
        \end{equation}
        $B \noinfo A$ means that, for all choices of $\sigma^{ABC}$ and
        the $B$-maps $\mathcal{F}_{k}$, $\rho^{A}_0 = \rho^{A}_1$.
\end{quote}

We can give a heuristic summary of the three conditions as follows.
Locality (I) says that ignorance (about $B$) doesn't hurt.
Locality (II) says that knowledge (of the state of $B$) doesn't help.
Locality (III) says that nothing we can do (to $B$) will make any
difference.  In fact, as we will now show, these three conditions
are completely equivalent, so any of them may be used as the definition
for the locality of the dynamical evolution of $C$ with context $B$.

Locality (III) clearly implies Locality (II), since
the input state $\sigma^{ABC}$ could
possibly be a product pure state, and the operations $\mathcal{F}_{k}$
could simply reset the state of $B$ to given fixed states $\ket{k}$.
Locality (I) also implies Locality (III).  Given trace-preserving
maps $\mathcal{F}_k$ on $B$ states, we can define
\begin{equation}
    \sigma_{k}^{ABC} = \mathcal{F}_k \otimes \unity^{AC}
        \left ( \sigma^{ABC} \right ) .
\end{equation}
From this we can see that $\partialtr{B} \sigma_{k}^{ABC} = \sigma^{AC}$,
the same state for every choice of $k$.
By Locality (I),
\begin{equation}
    \partialtr{BC} \cpmap{ABC}{} \left (  \sigma_{k}^{ABC} \right )
    = \cpmap{A}{AC} \left (  \partialtr{B} \sigma_{k}^{ABC} \right )
    = \cpmap{A}{AC} \left ( \sigma^{AC} \right ) ,
\end{equation}
which is manifestly independent of $k$, and so Locality (III) holds.
To show that all three conditions are independent, therefore, we
need to prove that Locality (II) implies Locality (I).

For a given system, we can find an operator basis of pure states,
so that any operator $X$ can be written as a linear combination of
projections:
\begin{equation}
    X = \sum_{n} c_{n} \proj{n} .
\end{equation}
If the underlying Hilbert space has dimension $d$, then the set of
pure states $\{ \ket{n} \}$ will have $d^2$ elements.  (It follows
that the vectors $\ket{n}$ cannot form an orthogonal set.)  Suppose
we choose states $\ket{\alpha}$ to yield an operator basis for $A$,
$\ket{\beta}$ to yield an operator basis for $B$, and $\ket{\gamma}$
to yield an operator basis for $C$.  Then the product states
$\ket{\alpha} \otimes \ket{\beta} \otimes \ket{\gamma}$ will yield
an operator basis for the composite system $ABC$.  This means that
any density operator $\sigma^{ABC}$ can be written
\begin{equation}
    \sigma^{ABC} = \sum_{\alpha \beta \gamma} s_{\alpha \beta \gamma}
        \proj{\alpha} \otimes \proj{\beta} \otimes \proj{\gamma} .
\end{equation}
If we take a partial trace over $B$, then the subsystem state
$\sigma^{AC}$ is written
\begin{equation}
    \sigma^{AC} =  \sum_{\alpha \gamma} s_{\alpha \gamma}
        \proj{\alpha} \otimes \proj{\gamma}
\end{equation}
where $s_{\alpha \gamma} = \bigsum{\beta} s_{\alpha \beta \gamma}$.

Suppose Locality (II) holds for the evolution $\cpmap{ABC}{}$.
We wish to construct the map $\cpmap{A}{AC}$ that takes initial
$AC$-states to final $A$-states.  Fix a particular $B$-state
$\ket{0}$ (which should be one of the states $\ket{\beta}$ that
give the operator basis), and define
\begin{equation}
    \cpmap{A}{AC} \left ( \sigma^{AC} \right ) =
        \partialtr{BC} \cpmap{ABC}{} \left (
        \proj{0} \otimes \sigma^{AC} \right ) .
\end{equation}
This is by construction a trace-preserving CP map.  Now, for
any states $\ket{\alpha}$, $\ket{\beta}$ and $\ket{\gamma}$,
Locality (II) implies that
\begin{equation}
    \partialtr{BC} \cpmap{ABC}{}
    \left ( \proj{\alpha} \otimes \proj{\beta} \otimes \proj{\gamma} \right )
    = \partialtr{BC} \cpmap{ABC}{}
    \left ( \proj{\alpha} \otimes \proj{0} \otimes \proj{\gamma} \right ) .
\end{equation}
Therefore, for any initial $ABC$-state $\sigma^{ABC}$,
\begin{eqnarray}
    \partialtr{BC} \cpmap{ABC}{} \left ( \sigma^{ABC} \right )
    & = &
    \sum_{\alpha \beta \gamma} s_{\alpha \beta \gamma} \partialtr{BC} \cpmap{ABC}{}
    \left ( \proj{\alpha} \otimes \proj{\beta} \otimes \proj{\gamma} \right )
        \nonumber \\
    & = &
    \sum_{\alpha \beta \gamma} s_{\alpha \beta \gamma} \partialtr{BC} \cpmap{ABC}{}
    \left ( \proj{\alpha} \otimes \proj{0} \otimes \proj{\gamma} \right )
        \nonumber \\
    & = &
    \sum_{\alpha \gamma} s_{\alpha \gamma} \partialtr{BC} \cpmap{ABC}{}
    \left ( \proj{\alpha} \otimes \proj{0} \otimes \proj{\gamma} \right )
        \nonumber \\
    & = &
    \partialtr{BC} \cpmap{ABC}{} \left ( \proj{0} \otimes \sigma^{AC} \right )
        \nonumber \\
    & = & \cpmap{A}{AC} \left ( \sigma^{AC} \right ) .
\end{eqnarray}
The map $\cpmap{A}{AC}$ thus satisfies the requirement of
Locality (I).  The three conditions are all equivalent, as promised.
Each of them captures the notion of the locality of the dynamical
evolution of $A$ with context $C$.

Suppose that an independent system $R$ is appended to $ABC$, so that the
overall system evolves according to $\cpmap{ABC}{} \otimes \unity^{R}$,
where $\unity^{R}$ is the identity map.  Then if $B \noinfo A$, a
straightforward derivation using Locality (I) shows that $B \noinfo AR$
and $BR \noinfo A$.  Note that this is a statement about the CP maps and
remains true even if the initial quantum state has entanglement between
$R$ and $ABC$.

\section{Precursor subspaces}

Our next task is to explore some of the implications of locality
in the evolution of quantum systems.  To do this, we will find
it convenient (as we will see in the next section)
to introduce the idea of a {\em precursor subspace}.

Let $\cpmap{}{}$ be a trace-preserving CP map on density operators.
(We make no assumption about the input and output states
of $\cpmap{}{}$; these may be states of the same system, or of different
systems.)  It may happen that $\cpmap{}{}$ takes a pure input state to
a pure output state:
\begin{equation}
   \cpmap{}{} \left ( \proj{\phi} \right ) = \proj{\psi}  .
\end{equation}
In this case, we say that $\ket{\phi}$ is a {\em precursor} of
$\ket{\psi}$ under $\cpmap{}{}$.  In this section we make
some observations about pure states and their precursors.

Suppose the operators $A_{\mu}$ give an operator sum
representation for $\cpmap{}{}$.
If $\ket{\phi}$ is a precursor of $\ket{\psi}$ under $\cpmap{}{}$,
then for all $\mu$,
\begin{equation}
    A_{\mu} \ket{\phi} = a_{\mu} \ket{\psi}  \label{opsumfact-eq}
\end{equation}
where the $a_{\mu}$'s are scalars.  To see this, let
$\ket{\hat{\psi}_{\mu}} = A_{\mu} \ket{\phi}$.  (The
``hat'' reminds us that this vector will not in general
be normalized, even if $\ket{\phi}$ is.)  Then
\begin{eqnarray}
    \proj{\psi}
        & = & \cpmap{}{} \left ( \proj{\phi} \right ) \\
        & = & \sum_{\mu} A_{\mu} \proj{\phi} {A_{\mu}}^{\dagger} \\
        & = & \sum_{\mu} \proj{\hat{\psi}_{\mu}} .
\end{eqnarray}
The only way that the positive operators $\proj{\hat{\psi}_{\mu}}$ could
sum to the rank-1 projection $\proj{\psi}$ would be if each of them were
multiples of $\proj{\psi}$.  This means that $\ket{\hat{\psi}_{\mu}} =
a_{\mu} \ket{\psi}$ for every $\mu$.

The converse of this is also true.  If $A_{\mu} \ket{\phi} = a_{\mu} \ket{\psi}$
for all $\mu$, then $\cpmap{}{} \left ( \proj{\phi} \right ) = \proj{\psi}$.
(The only issue here is normalization, which follows from the fact that
$\cpmap{}{}$ is trace-preserving.)

For a state vector $\ket{\psi}$ in the output space, we define
\begin{equation}
    \subspace{\psi} = \left \{ \ket{\hat{\phi}} \, : \,
        \cpmap{}{} \left ( \proj{\hat{\phi}} \right )
        = \lambda \proj{\psi} \, , \, \lambda \geq 0 \right \} .
\end{equation}
This is the set of input vectors
which are (up to normalization) precursors of $\ket{\psi}$.
This set $\subspace{\psi}$ is a subspace, as can be seen from
the previous fact.  Pick an operator
sum representation for $\cpmap{}{}$ given by operators $A_{\mu}$.
If $\ket{\phi}$ and $\ket{\phi'}$ are in $\subspace{\psi}$,
then
\begin{equation}
   A_{\mu} \left ( \alpha \ket{\phi} + \alpha' \ket{\phi'} \right )
        = a_{\mu} \alpha \ket{\psi} + a_{\mu}' \alpha' \ket{\psi} =
        b_{\mu} \ket{\psi} .
\end{equation}
This means that $\cpmap{}{}$ will take the superposition of $\ket{\phi}$
and $\ket{\phi'}$ to a multiple of $\ket{\psi}$, and so the
superposition lies in $\subspace{\psi}$.  The set $\subspace{\psi}$
is therefore a subspace.
We call this the {\em precursor subspace} of $\ket{\psi}$.
Notice that, even though the map $\cpmap{}{}$ acts on operators,
the precursor subspace exists in the underlying Hilbert space.

Given a map $\cpmap{}{}$ and
any $\ket{\psi}$, there is a precursor subspace $\subspace{\psi}$.
However, it may be the case that this subspace is null.  For example,
suppose we have a qubit whose pure states are spanned by computational
basis states $\ket{0}$ and $\ket{1}$.  Consider the map $\cpmap{}{}$ which
takes every input state $\sigma$ to $\cpmap{}{}(\sigma) = \proj{0}$.  Then the
precursor space of $\ket{0}$ is the whole Hilbert space for the qubit,
but the precursor space of any other state will be null.

How are the precursor subspaces for two distinct pure states related
to each other?  It is easy to see that the two precursor subspaces can
only intersect in the null space.
Now suppose that $\ket{\phi}$ and $\ket{\phi'}$ are precursors for
$\ket{\psi}$ and $\ket{\psi'}$, respectively.
Since fidelity is monotonic under CP maps \cite{mikeandike},
\begin{equation}
    \left | \amp{\phi}{\phi'} \right |^{2} \leq
        \left | \amp{\psi}{\psi'} \right |^{2} .
\end{equation}
As a corollary, if $\ket{\psi}$ and $\ket{\psi'}$ are
orthogonal, their precursors must also be orthogonal.
The precursor subspaces for orthogonal states are orthogonal
subspaces.

\section{Autonomy}

Suppose quantum system $A$ is described by a Hilbert space
$\hilbert^{A}$ of dimension $d_{A}$.  Every trace-preserving
CP map $\cpmap{}{}$ on $A$ has a unitary representation---in fact,
many different unitary representations, employing environment
systems of various sizes.
However, any CP map on $A$ states can be represented
using an environment system $E$ whose Hilbert space
dimension is no larger than $d_{A}^2$.  We can classify
the maps by their {\em rank}, the Hilbert space dimension of the
smallest environment needed to give a unitary representation.  For
any $\cpmap{}{}$,
\begin{equation}
    1 \leq \rank \cpmap{}{} \leq d_{A}^2 .
\end{equation}
(The rank of $\cpmap{}{}$ is also the minimum number of operators
required for an operator-sum representation of $\cpmap{}{}$.)
The minimal-rank operations are those which require no environment
system at all---that is, the maps that are already unitary.

What are the minimal-rank operations in the case where the input
and output states belong to different systems?
Consider a CP map $\cpmap{A}{AC}$ that takes
states of a composite system $AC$ to states of its subsystem $A$.
Systems $A$ and $C$ are described by Hilbert spaces
of dimension $d_A$ and $d_C$, respectively, and the
Hilbert space for $AC$ has dimension $d_A d_C$.  The
minimal-rank operations of this type are those that
do not require an external environment for their unitary
representation.  We call this property {\em autonomy}:
\begin{quote}
{\bf Autonomy.}  The map $\cpmap{A}{AC}$ is {\em autonomous}
    if there exists a unitary operator $U$ on $AC$ such that
    \begin{equation}
        \cpmap{A}{AC} (\sigma) = \partialtr{C} U \sigma U^{\dagger} .
    \end{equation}
    for any $AC$-state $\sigma$.  In other words, a unitary
    representation for an autonomous CP map does not require
    the introduction of any additional environment system.
\end{quote}

It will turn out that autonomy is equivalent to two other
technical conditions on the map $\cpmap{A}{AC}$, which are:
\begin{quote}
{\bf Uniform dimension condition (UDC).}  The map $\cpmap{A}{AC}$
    satisfies the {\em uniform dimension condition} if, for any
    $\ket{\psi} \in \hilbert^{A}$, $\dim \subspace{\psi} = d_{C}$.

{\bf Output rank condition (ORC).}  Suppose we add an
    ancilla system $R$ to $AC$ and prepare the overall system
    in an initial state $\ket{\Psi}$ in which $R$ is maximally
    entangled with $AC$.  The entire system evolves according to
    the map $\unity^{R} \otimes \cpmap{A}{AC}$, leading
    to a final state $\rho$ of $RA$.
    We say that $\cpmap{A}{AC}$ satisfies the
    {\em output rank condition} if $\rho$ has rank $d_{C}$.
\end{quote}

To show that these three are equivalent, we will prove that
autonomy implies the UDC, the ORC implies autonomy, and the UDC
implies the ORC.

{\bf Autonomy $\Rightarrow$ UDC.}  Suppose $\cpmap{A}{AC}$ is autonomous,
with $U$ being the implied unitary operator on $AC$.  Let
$\ket{\psi} \in \hilbert^{A}$.  Then
\begin{equation}
    \subspace{\psi} = \left \{ U^{\dagger} \left (
            \ket{\gamma} \otimes \ket{\psi} \right )
            \, : \,  \ket{\gamma} \in \hilbert^{C} \right \} .
\end{equation}
This clearly has dimension $d_{C}$.

{\bf ORC $\Rightarrow$ Autonomy.}  Suppose we add the ancilla system
$R$, and start with the maximally entangled state $\ket{\Psi}$ of $RAC$,
which maps under $\unity^{R} \otimes \cpmap{A}{AC}$ to the density operator
$\rho^{RA}$.  Also suppose that $\rank \rho = d_{C}$.
Then we can purify the final state $\rho^{RA}$ by appending
a system of dimension $d_{C}$---in
particular, by appending $C$ itself.
This yields a pure state $\ket{\Psi'}$ such that
\begin{equation}
    \rho^{RA} = \partialtr{C} \proj{\Psi'} .
\end{equation}
The state of $R$ alone has not changed under the evolution by
$\unity \otimes \cpmap{A}{AC}$.  Both $\ket{\Psi}$ and $\ket{\Psi'}$
are purifications of the same state of $R$, and hence are related by some
unitary operator $U^{AC}$ on $\hilbert^{AC}$.  Thus,
\begin{equation}
    \ket{\Psi'} = \left ( 1^{R} \otimes U^{AC} \right ) \ket{\Psi} .
\end{equation}
The unitary operator $U^{AC}$, together with the partial trace over $C$,
defines an autonomous CP map from $AC$ states to $A$ states.
But such a map is completely specified by its action on a single
maximally entangled input state of $RAC$, namely $\ket{\Psi}$.
Thus, this map must be the same as $\cpmap{A}{AC}$ itself,
and so $\cpmap{A}{AC}$ is autonomous.

{\bf UDC $\Rightarrow$ ORC.}  It remains to show that the uniform
dimension condition implies the output rank condition.  The ORC
states that, for an input state $\ket{\Psi}$ that is maximally
entangled between $R$ and $AC$, the output state $\rho^{RA} =
\unity^{R} \otimes \cpmap{A}{AC} \left ( \proj{\Psi} \right )$
has rank $d_{C}$.

In fact, without any assumptions about $\cpmap{A}{AC}$,
we can show that this output state has rank at least $d_{C}$.
We can write
\begin{equation}
    \rho^{RA} = \sum_{\alpha } P_{\alpha}
        \proj{\Phi_{\alpha}},
\end{equation}
where $\alpha$ runs from 1 to $\rank \rho^{RA}$, and
the states $\ket{\Phi_{\alpha}}$ are pure entangled states
of $RA$.  If we let
\begin{equation}
    \begin{array}{rcl}
    \rho_{\alpha}^{R} & = & \partialtr{A} \proj{\Phi_{\alpha}} \\
    \rho_{\alpha}^{A} & = & \partialtr{R} \proj{\Phi_{\alpha}}
    \end{array}
\end{equation}
then $\rank \rho_{\alpha}^{R} = \rank \rho_{\alpha}^{A} \leq d_{A}$.
Since
\begin{equation}
    \rho^{R} = \sum_{\alpha} P_{\alpha} \rho_{\alpha}^{R} ,
\end{equation}
it follows that $\rank \rho^{R} \leq ( \rank \rho^{RA} )  d_{A}$.
In our case, the input state $\ket{\Psi}$ is maximally entangled
between $R$ and $AC$ and the system $R$ evolves according to the
identity map $\unity^{R}$, so that $\rank \rho^{R} = d_{A} d_{C}$.
Therefore, $\rank \rho^{RA} \geq d_{C}$.

Now we show that if $\cpmap{A}{AC}$ satisfies the uniform dimension
condition, then $\rank \rho^{RA} \leq d_{C}$ as well.
This will require a much lengthier proof.  Our argument
is based on the following general fact.
Suppose $\cpmap{}{}$ is a CP map with operator-sum representation
$\left \{ A_{\mu} \right \}$, and suppose we have vectors
$\ket{\alpha_{1}} , \ldots , \ket{\alpha_{n}}$ which are precursor
states to pure states:
\begin{equation}
    \cpmap{}{} \left ( \proj{\alpha_{k}} \right ) = \proj{\phi_{k}}
\end{equation}
for $k = 1 , \ldots , n$.  Let $\ket{\phi} = \bigsum{k} c_{k}
\ket{\alpha_{k}}$.  Then the density operator
\begin{equation}
    \rho = \cpmap{}{} \left ( \proj{\phi} \right )
\end{equation}
has rank no larger than $n$.  To see this, note first that
\begin{equation}
    A_{\mu} \ket{\alpha_{k}} = \beta_{k \mu} \ket{\phi_{k}} .
\end{equation}
Then
\begin{equation}
    \rho = \sum_{k,k'} \left (
        \sum_{\mu} \beta_{k \mu} \beta_{k' \mu}^{\ast} c_{k} c_{k'}^{\ast}
        \right )  \ket{\phi_{k}} \bra{\phi_{k'}} .
\end{equation}
This operator obviously has support contained in the subspace spanned
by the image states $\ket{\phi_{k}}$, which has dimension no larger
than $n$.  Thus $\mbox{rank} \, \, \rho \leq n$.

Our plan is to write a maximally entangled input state of $RAC$
as a superposition of $d_{C}$ states that are precursors of pure
states under $\unity^{R} \otimes \cpmap{A}{AC}$.  It will follow
that $\rank \rho^{RA} \leq d_{C}$.

Now for the details.  Suppose that $\cpmap{A}{AC}$ satisfies the UDC.
Pick an orthonormal basis $\left \{ \ket{k} \right \}$ for $\hilbert^{A}$.
For each $\ket{k}$, we have a precursor subspace $\subspace{k}$.
There are $d_{A}$ such subspaces, and they must be orthogonal to
each other (since otherwise two non-orthogonal input states could map
to orthogonal output states).

Assuming the uniform dimension condition,
each of the precursor subspaces has dimension $d_{C}$.
We will now construct another subspace $\mathcal{T}_{1}$ of
dimension $d_{A}$ that ``cuts across'' the precursor
subspaces in a special way.  To begin with, we note that
any vector $\ket{\phi} \in \hilbert^{AC}$ can be written
$\ket{\phi} = \bigsum{k} a_{k} \ket{\phi_{k}}$, where
$\ket{\phi_{k}} \in \subspace{k}$.  Furthermore, for a
given $\ket{\phi}$, the (normalized) vectors $\ket{\phi_{k}}$
are unique up to phase.

Now pick a particular $\ket{\psi} \in \hilbert^{A}$ so that
$\ket{\psi} = \bigsum{k} c_{k} \ket{k}$ with $c_{k} \neq 0$.
Let $\ket{\phi}$ be some precursor of $\ket{\psi}$.
We can write the precursor state $\ket{\phi}$ as a superposition
of states in the subspaces $\subspace{k}$:
    \begin{equation}
        \ket{\phi} = \sum_{k} a_{k} \ket{\phi_{k}} .
    \end{equation}
Since the magnitudes of inner products cannot decrease under $\cpmap{}{}$,
we know that $| a_{k} |^2 \geq | c_{k} |^2$.  But since both
$\ket{\psi}$ and its precursor $\ket{\phi}$
are normalized,
\begin{equation}
    \sum_{k} | a_{k} |^2 = \sum_{k} | c_{k} |^2 = 1 .
\end{equation}
Therefore, $| a_{k} |^2 = | c_{k} |^2$ for all values of $k$.
By adjusting the phases of the $\ket{\phi_{k}}$ basis precursor states,
we can arrange for $a_{k} = c_{k}$.
Once we have done this, the precursor of
$\ket{\psi} = \bigsum{k} c_{k} \ket{k}$ will be
\begin{equation}
    \ket{\phi} = \sum_{k} c_{k} \ket{\phi_{k}} .
\end{equation}

Our subspace $\mathcal{T}_{1}$ is the subspace spanned by the basis
precursor states $\ket{\phi_{k}}$ that we have chosen.
There are $d_{A}$ of these, so that is the dimension of $\mathcal{T}_{1}$.
Our next step is to show that {\em any} pure state in $\mathcal{T}_{1}$
is a precursor of some pure state in $\hilbert^{A}$.  Introduce an
operator-sum representation for the map $\cpmap{A}{AC}$, given
by operators $A_{\mu}$.  (The operators $A_{\mu}$ act on vectors
in $\hilbert^{AC}$ and map them to vectors in $\hilbert^{A}$.)
From Equation~\ref{opsumfact-eq}, we see that
\begin{eqnarray}
    A_{\mu} \ket{\phi} & = & \alpha_{\mu} \ket{\psi} \nonumber \\
    A_{\mu} \ket{\phi_{k}} & = & \beta_{\mu k} \ket{k}
\end{eqnarray}
for some scalars $\alpha_{\mu}$ and $\beta_{\mu k}$.  Writing
$\ket{\phi}$ in terms of the basis precursors $\ket{\phi_{k}}$
and $\ket{\psi}$ in terms of the basis states $\ket{k}$, we
obtain
\begin{equation}
    \sum_{k} \alpha_{\mu} c_{k} \ket{k} = \sum_{k} \beta_{\mu k} c_{k} \ket{k} .
\end{equation}
This implies that $\alpha_{\mu} = \beta_{\mu k}$ for all values
of $k$ and $\mu$.

Now consider another vector $\ket{\phi'}$ in $\mathcal{T}_{1}$, which is
a superposition of our basis precursor states:
\begin{equation}
    \ket{\phi'} = \sum_{k} c_{k}' \ket{\phi_{k}} .
\end{equation}
The operators of the operator-sum representation act on this
vector to yield
\begin{equation}
    A_{\mu} \ket{\phi'} = \alpha_{\mu} \sum_{k} c_{k}' \ket{k}
            =   \alpha_{\mu} \ket{\psi'} ,
\end{equation}
where $\ket{\psi'} = \bigsum{k} c_{k}' \ket{k}$.  This in turn
implies that
\begin{equation}
    \cpmap{A}{AC} \left ( \proj{\phi'} \right ) = \proj{\psi'} .
\end{equation}

To sum up, we have found a subspace $\mathcal{T}_{1}$ such that
every vector in $\hilbert^{A}$ has a precursor in $\mathcal{T}_{1}$,
every vector in $\mathcal{T}_{1}$ is the precursor of some vector
in $\hilbert^{A}$, and the relation between precursor and image
is linear.  Furthermore, the intersection of $\mathcal{T}_{1}$
with any precursor subspace $\subspace{\psi}$ is one-dimensional.
We let $\ket{\Upsilon_{1k}} = \ket{\phi_{k}}$; the vectors
$\ket{\Upsilon_{1k}}$ form a basis for $\mathcal{T}_{1}$.

Now we turn our attention to $\mathcal{T}_{1}^{\perp}$, which is
a subspace of $\hilbert^{AC}$ of dimension $(d_{C} - 1) d_{A}$.
Every state $\ket{\psi}$ in $\hilbert^{A}$ will have a precursor
subspace within $\mathcal{T}_{1}^{\perp}$ of dimension $d_{C} - 1$.
Therefore, $\mathcal{T}_{1}^{\perp}$ satisfies a uniform
dimension condition with a reduced precursor subspace dimension
$d_{C} - 1$.  This in turn means that we can repeat our process to arrive
at a new subspace $\mathcal{T}_{2}$ orthogonal to $\mathcal{T}_{1}$
such that every vector in $\mathcal{T}_{2}$ is a precursor of some
pure state and the relation between precursor and image is linear.
Also, the intersection of $\mathcal{T}_{2}$ and any precursor
subspace $\subspace{\psi}$ will be one-dimensional.  We let the
vectors $\ket{\Upsilon_{2k}}$ in $\mathcal{T}_{2}$
be the precursors of the basis vectors $\ket{k}$.

We can generalize this process.  At the $n$th stage, we find the
subspace $\mathcal{T}_{n}$ that is perpendicular to the linear
span of $\mathcal{T}_{1}$ through $\mathcal{T}_{n-1}$.  The new
subspace $\mathcal{T}_{n}$ has dimension $d_{A}$, and
each of its elements is a precursor of some state in $\hilbert^{A}$.
The relation between precursor in $\mathcal{T}_{n}$ and
image in $\hilbert^{A}$ is linear.
Every precursor subspace $\subspace{\psi}$ in $\hilbert^{AC}$ has
a one-dimensional intersection with $\mathcal{T}_{n}$.
Finally, we identify basis vectors $\ket{\Upsilon_{nk}}$ that
are precursors of basis vectors $\ket{k}$.

Now introduce the ancilla system $R$ and let the whole system
evolve according to the map $\unity^{R} \otimes \cpmap{A}{AC}$.
Imagine that we have an input pure state
\begin{equation}
    \ket{\Psi_{n}} = \sum_{k} c_{k} \ket{\alpha_{nk}^{R}}
        \otimes \ket{\Upsilon_{nk}} .
\end{equation}
In other words, the input state $\ket{\Psi_{n}}$ is an entangled
state whose support in $\hilbert^{AC}$ is entirely contained in
$\mathcal{T}_{n}$.  By our construction of the subspace
$\mathcal{T}_{n}$,
\begin{equation}
    \unity^{R} \otimes \cpmap{A}{AC} \left ( \proj{\Psi_{n}} \right )
        = \proj{\Phi_{n}}  \label{pureoutput-eq}
\end{equation}
where
\begin{equation}
    \ket{\Phi_{n}} = \sum_{k} c_{k} \ket{\alpha_{nk}^{R}} \otimes \ket{k} .
\end{equation}
A pure entangled state that is supported within
$\mathcal{T}_{n}$ maps to a pure entangled output state.

Now, any entangled input state $\ket{\Psi}$ of $RAC$ can be written
as
\begin{equation}
    \ket{\Psi} = \sum_{n} a_{n} \ket{\Psi_{n}} ,  \label{entangledinput-eq}
\end{equation}
where the $\ket{\Psi_{n}}$ states have $AC$ support in $\mathcal{T}_{n}$.
That is, $\ket{\Psi}$ is a superposition of $d_{B}$ states each of which
is a precursor of some pure state under $\unity^{R} \otimes \cpmap{A}{AC}$.
The output state
\begin{equation}
    \rho^{RA} = \unity^{R} \otimes \cpmap{A}{AC} \left ( \proj{\Psi} \right )
\end{equation}
therefore has a rank no larger than $d_{C}$, as we wished to prove.

We have now shown that $\rank \rho^{RA} \geq d_{C}$ and
$\rank \rho^{RA} \leq d_{C}$, so $\rank \rho^{RA} = d_{C}$, and
the output rank condition (ORC) holds for $\cpmap{A}{AC}$.
Autonomy, the UDC and the ORC are all equivalent conditions on $\cpmap{A}{AC}$.

Autonomy will prove to be a useful idea when considering locality in systems
which have unitary global evolution.  This is the subject of the next section.

\section{Global unitarity}

Return to the situation in which the joint system $ABC$ evolves according to
$\cpmap{ABC}{}$ such that $B \noinfo A$.  This implies the existence of a local
CP map $\cpmap{A}{AC}$.  What can we say about the local evolution $\cpmap{A}{AC}$
if we know that the global evolution $\cpmap{ABC}{}$ is in fact unitary?

Let $\ket{\psi}$ be a pure output state of $A$.  We wish to consider
$\ket{\psi}$ as an output of each of the maps $\cpmap{A}{AC}$ and
$\cpmap{A}{ABC}$. Let $\subspace{\psi}^{AC}$ and $\subspace{\psi}^{ABC}$
be the precursor subspaces of $\ket{\psi}$ for these two maps.  If
$\ket{\phi} \in \subspace{\psi}^{AC}$, then every vector of the form
$\ket{\beta} \otimes \ket{\phi}$ (where $\ket{\beta}$ is a $B$ state)
must be in $\subspace{\psi}^{ABC}$.  This implies that
$\hilbert^{B} \otimes \subspace{\psi}^{AC} \subseteq \subspace{\psi}^{ABC}$.

Conversely, suppose that $\ket{\Phi} \in \subspace{\psi}^{ABC}$.  This state
may have entanglement between $B$ and $AC$, but in any case we can write it as
\begin{equation}
    \ket{\Phi} = \sum_{k} \alpha_{k} \ket{k} \otimes \ket{\phi_{k}}
\end{equation}
where the $\ket{k}$ states are an orthonormal basis for $\hilbert^{B}$.
The partial trace over $B$ of this state yields the mixed $AC$ state
\begin{equation}
    \sigma^{AC} = \sum_{k} \left | \alpha_{k} \right |^{2} \proj{\phi_{k}} .
\end{equation}
We know that $\cpmap{A}{AC} \left ( \sigma^{BC} \right ) = \proj{\psi}$.
Therefore, it must be that $\ket{\phi_{k}} \in \subspace{\psi}^{AC}$ for
every $k$ with $\alpha_{k} \neq 0$, and
\begin{equation}
    \ket{\Phi} \in \hilbert^{B} \otimes \subspace{\psi}^{AC} .
\end{equation}
Thus $\subspace{\psi}^{ABC} = \hilbert^{B} \otimes \subspace{\psi}^{AC}$.

The map $\cpmap{A}{ABC}$ is clearly autonomous, and so satisfies the uniform
dimension condition.  The precursor subspace $\subspace{\psi}^{ABC}$ has
dimension $d_{BC} = d_{B} d_{C}$.  Since this subspace is a tensor product of
$\hilbert^{B}$ and $\subspace{\psi}^{AC}$, we find that $\dim \subspace{\psi}^{AC}
= d_{C}$.  Because this is independent of the choice of $\ket{\psi}$, we
see that $\cpmap{A}{AC}$ satisfies the uniform dimension condition.  In short,
if $\cpmap{ABC}{}$ is unitary and $B \noinfo A$, then $\cpmap{A}{AC}$ must be
autonomous.

Furthermore, we can show that such a unitary map decomposes in an elegant way.
Let $U$ be the unitary operator acting on $\hilbert^{ABC}$ that gives rise to
the unitary map $\cpmap{ABC}{}$.  We append ancilla systems $R_{B}$ and $R_{AC}$
and prepare the initial state so that $B$ and $R_{B}$ are maximally entangled,
as are $AC$ and $R_{AC}$.  (This of course means that $R_{B}R_{AC}$ is maximally
entangled with $ABC$.)  Call this input state $\ket{\Psi}$.

Now we construct two scenarios.  In the first, our composite system
evolves according to the unitary operator $U$.  (Technically, for
the entire system $ABC R_{B} R_{AC}$ this would be the operator
$U^{ABC} \otimes 1^{R_{B}} \otimes 1^{R_{AC}}$.)  At the end,
the joint state is $\ket{\Phi_{U}}$.

\begin{figure}[htbp]
\begin{center}
\includegraphics[height=2.0in]{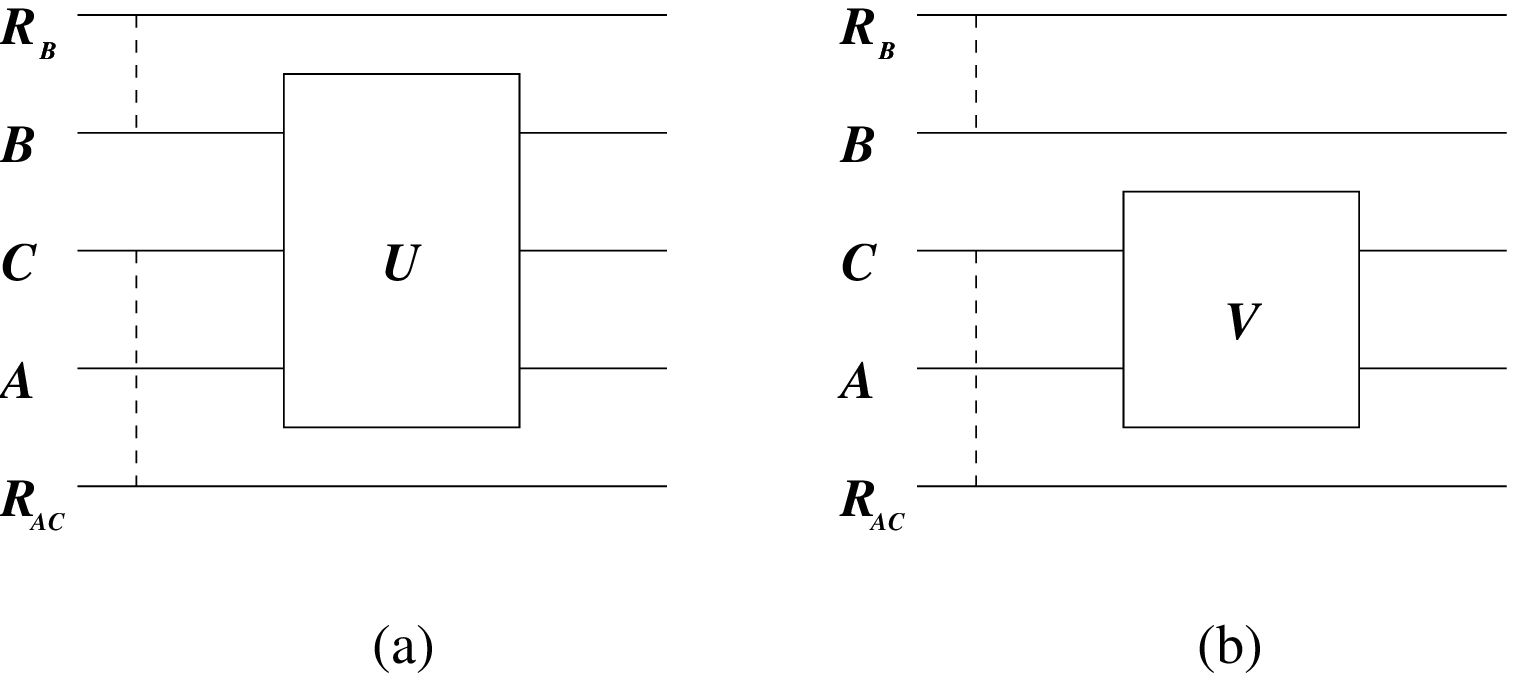}
\end{center}
\caption{Two scenarios for the evolution of the five-part system
    $ABC R_{B} R_{AC}$.  In (a), $ABC$ evolves by $U$, while in (b)
    only $AC$ evolves according to $V$.  Dotted lines indicate the
    initial entanglement of the system.
    \label{2scenarios-fig}}
\end{figure}

In the second scenario, we consider a unitary operator $V$
that gives a unitary representation of $\cpmap{A}{AC}$.
Since we have shown that $\cpmap{A}{AC}$ is autonomous,
$V$ is chosen only to act on $\hilbert^{AC}$.  That is,
for any $AC$ state $\sigma$,
\begin{equation}
   \cpmap{A}{AC} (\sigma) = \partialtr{B} V \sigma V^{\dagger} .
\end{equation}
The final state in this case is $\ket{\Phi_{V}}$.
The two scenarios are illustrated as circuit diagrams
in Figure~\ref{2scenarios-fig}.

\begin{figure}[htbp]
\begin{center}
\includegraphics[height=1.0in]{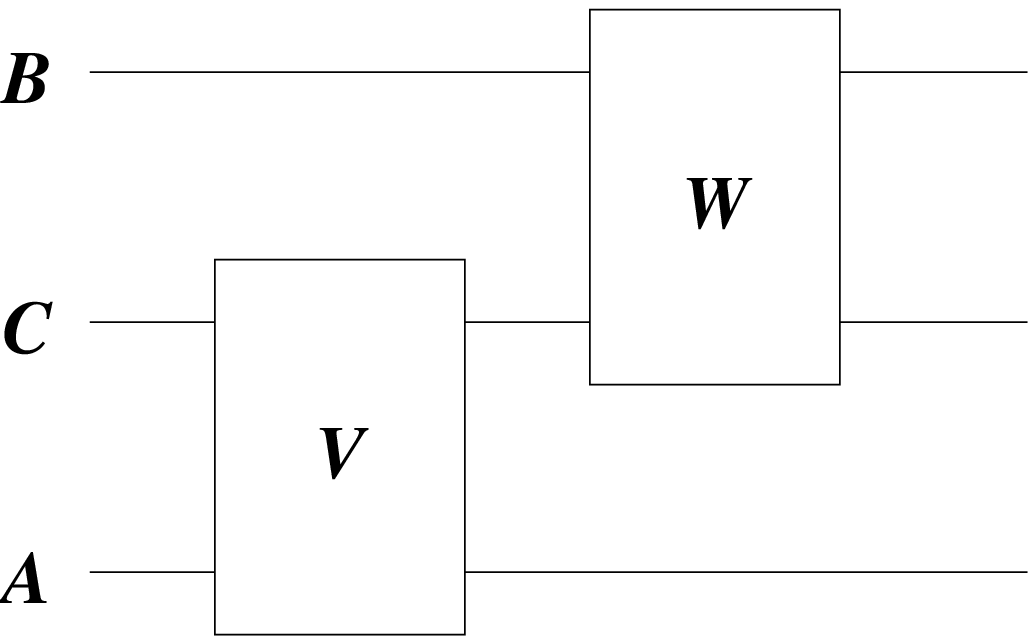}
\end{center}
\caption{Decomposition of a unitary operator for which $B \noinfo A$.
    \label{udecomp-fig}}
\end{figure}

How do the two final global states $\ket{\Phi_{U}}$ and $\ket{\Phi_{V}}$
differ from one another?  By the definition of the local map $\cpmap{A}{AC}$,
\begin{equation}
   \cpmap{A}{AC} \circ \partialtr{B} = \cpmap{A}{ABC}
     = \partialtr{BC} \circ \cpmap{ABC}{}
\end{equation}
(Equation~\ref{localityI-eq} above).  Thus for our extended system,
\begin{equation}
    \partialtr{BC}
      \circ \left ( \unity^{R_{B}} \otimes \cpmap{ABC}{} \otimes \unity^{R_{AC}} \right )
     = \left ( \cpmap{A}{AC} \otimes \unity^{R_{B}} \otimes \unity^{R_{AC}} \right )
        \circ \partialtr{B} .
\end{equation}
It follows that the final state of the subsystem $A R_{B} R_{AC}$ is
\begin{equation}
    \rho^{A R_{B}R_{AC}} = \partialtr{BC} \proj{\Phi_{U}} = \partialtr{BC} \proj{\Phi_{V}} .
\end{equation}
The states $\ket{\Phi_{U}}$ and $\ket{\Phi_{V}}$ are thus purifications of
the same marginal state on the subsystem $A R_{B}R_{AC}$.  This implies that
there is a unitary operator $W$ that acts only on the complementary system $BC$
such that
\begin{eqnarray}
    \ket{\Phi_{U}} & = &
        \left ( W^{BC} \otimes 1^{R_{B}} \otimes 1^{A R_{AC}} \right )
        \ket{\Phi_{V}} \nonumber \\
    \left ( U^{ABC} \otimes 1^{R_{B}} \otimes 1^{R_{AC}} \right ) \ket{\Psi}
        & = &
        \left ( W^{BC} \otimes 1^{R_{B}} \otimes 1^{A R_{AC}} \right )
            \nonumber  \\
        &   &  \mbox{\hspace{0.25in}} \times
        \left ( 1^{R_{B}B} \otimes V^{AC} \otimes 1^{R_{AC}} \right ) \ket{\Psi}
\end{eqnarray}
Since this is true for the input state $\ket{\Psi}$, which is maximally
entangled between $R_{B} R_{AC}$ and $ABC$, it follows that
\begin{equation}
    U^{ABC} = \left ( W^{BC} \otimes 1^{A} \right )
                \left ( 1^{B} \otimes V^{AC} \right ) .
\end{equation}
We have shown that any unitary map $\cpmap{ABC}{}$ for which $B \noinfo A$ can
be decomposed as shown in Figure~\ref{udecomp-fig}.
In this decomposition, systems $A$ and $C$ interact first, and then
systems $B$ and $C$ interact.  This causal structure clearly guarantees
that no information can be transferred from $B$ to $A$; we have now
shown that this sort of structure is the {\em only} way to guarantee
$B \noinfo A$ in a unitary map.

\begin{figure}[htbp]
\begin{center}
\includegraphics[height=1.0in]{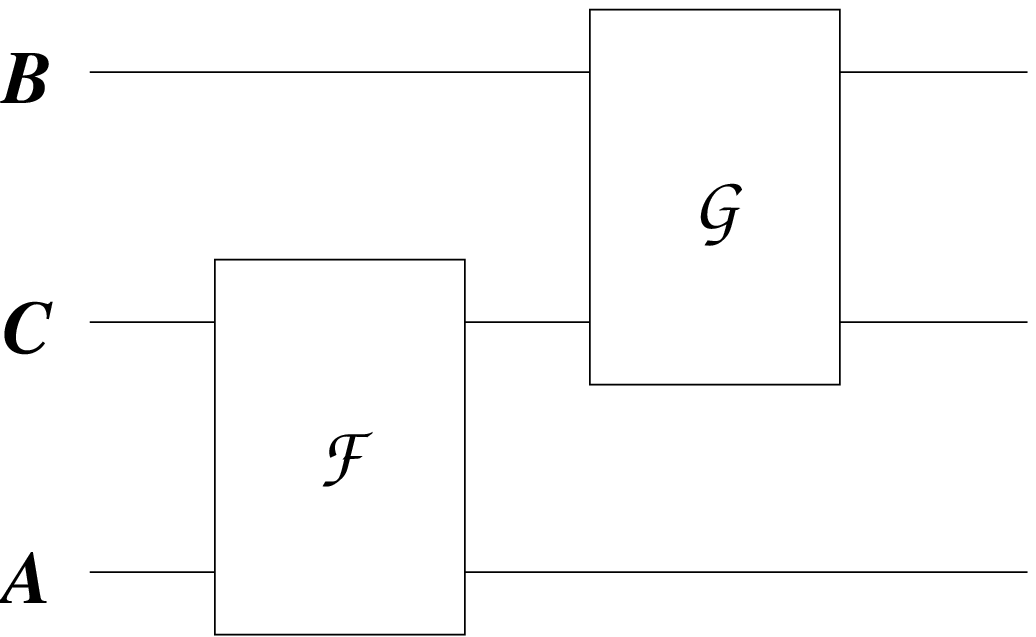}
\end{center}
\caption{Decomposition of an arbitrary CP map implying that $B \noinfo A$.
    \label{edecomp-fig}}
\end{figure}

The causal structure illustrated in Figure~\ref{udecomp-fig} could
apply to more general CP maps as well.  If an overall map $\cpmap{ABC}{}$
could be decomposed as shown in Figure~\ref{edecomp-fig}, then it would be
clearly true that $B \noinfo A$.  But does the converse hold?  If $B \noinfo A$
in this more general context, can we always decompose $\cpmap{ABC}{}$ as
shown in Figure~\ref{edecomp-fig}?

The answer is no.  It is easy to come up with a CP map on $ABC$ for which
$B \noinfo A$, but which cannot be written in this way.  Consider for
instance a map in which a measurement is performed on $A$, and its result
is written in the state of $B$ (erasing any previous state).  System $C$
evolves via the identity map.  This example cannot be decomposed in the
way suggested by Figure~\ref{edecomp-fig}, but clearly $B \noinfo A$.

On the other hand, we can find a unitary representation for any CP map.
If $B \noinfo A$, what can we say about the structure of such a representation?
First, let us consider the case of two systems $A$ and $B$ which evolve according
to a global CP map $\cpmap{AB}{}$, and for which $B \noinfo A$.  To construct
a unitary representation for $\cpmap{AB}{}$, we introduce an environment
system $E$ in a standard initial state $\ket{0}$.  Then there exists a
unitary operator $U$ on $ABE$ such that
\begin{equation}
    \cpmap{AB}{} \left ( \sigma^{AB} \right )
        = \partialtr{E} \left ( U \left ( \sigma^{AB} \otimes \proj{0} \right )
                U^{\dagger} \right )
\end{equation}
for any $\sigma^{AB}$.  Since $B \noinfo A$, there is a local map $\cpmap{A}{}$,
and this local map itself has a unitary representation.  Appending the environment
$E$ initially in $\ket{0}$, there is a unitary $V$ acting on $\hilbert^{AE}$ so
that
\begin{equation}
    \cpmap{A}{} \left ( \sigma^{A} \right )
        = \partialtr{E} \left ( V \left ( \sigma^{A} \otimes \proj{0} \right )
                V^{\dagger} \right )
\end{equation}
for all inputs $\sigma^{A}$.

\begin{figure}[htbp]
\begin{center}
\includegraphics[height=1.0in]{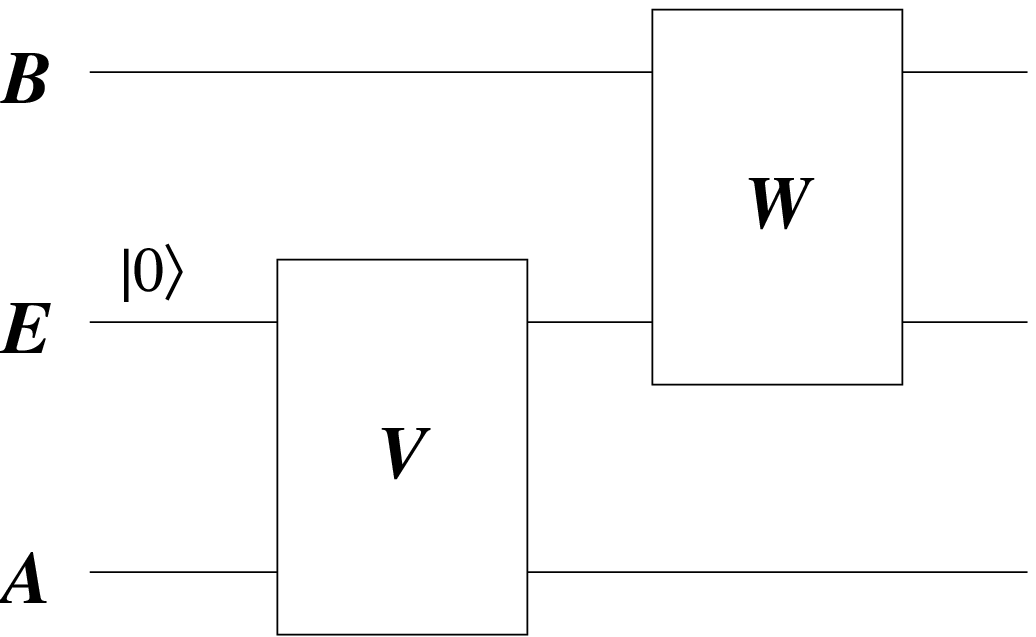}
\end{center}
\caption{Unitary representation for $\cpmap{AB}{}$ when $B \noinfo A$.
    \label{urepdecomp-fig}}
\end{figure}

Now append ancilla systems $R_{A}$ and $R_{B}$ which initially are maximally
entangled with $A$ and $B$, respectively.  The global initial state is $\ket{\Psi}$.
To this initial state we can apply either $U$ (to $ABE$) or $V$ (to $AE$ alone).
In either case, we will arrive at a final state that has the same marginal state
for $A R_{A} R_{B}$, and thus the two final states differ only by a unitary
transformation $W$ affecting only $B$ and $E$.  Pictorially, we have
Figure~\ref{urepdecomp-fig}.
This is not exactly the same as our previous result, since the input state of
the environment $E$ is fixed to be $\ket{0}$, unlike the system $C$ which can
have any input state.  We have nevertheless shown that $B \noinfo A$ implies that
the global map $\cpmap{AB}{}$ has a unitary representation in which the
environment interacts with $A$ and with $B$ sequentially.  Any information transfer
between the two systems is mediated by the system $E$, and this transfer can
only occur in one direction.

\begin{figure}[htbp]
\begin{center}
\includegraphics[height=1.35in]{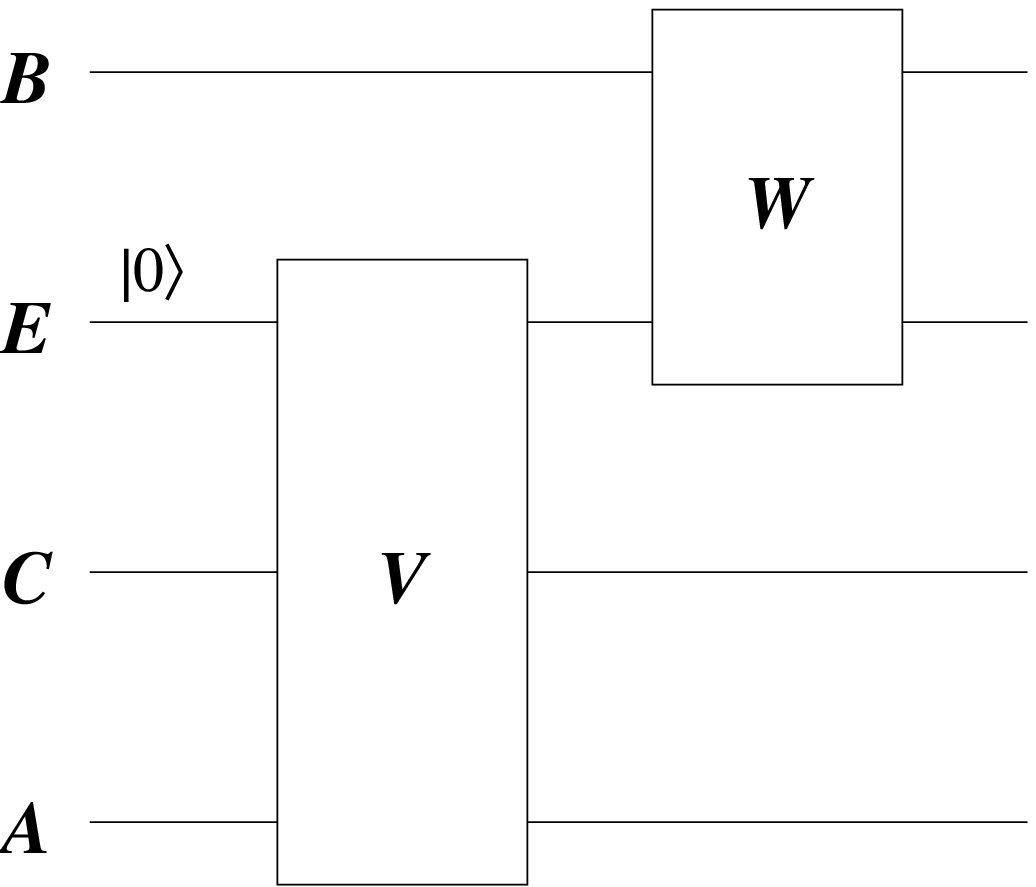}
\end{center}
\caption{Unitary representation for $\cpmap{ABC}{}$ with $B \noinfo A$.
    \label{3urepdecomp-fig}}
\end{figure}

\begin{figure}[htbp]
\begin{center}
\includegraphics[height=1.0in]{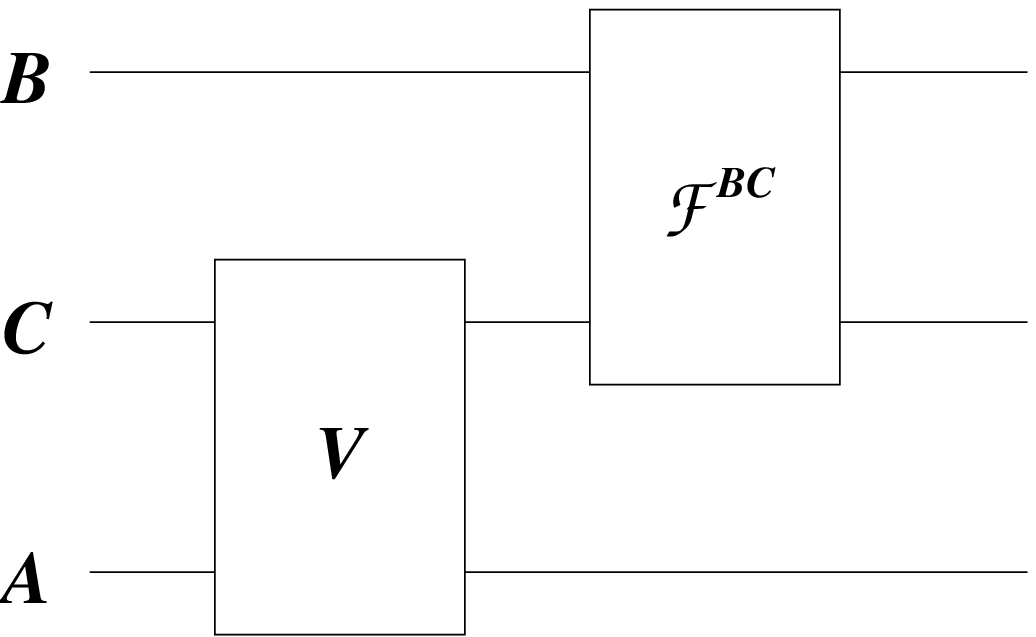}
\end{center}
\caption{If $\cpmap{A}{AC}$ is autonomous and $B \noinfo A$, then
    $\cpmap{ABC}{}$ can be decomposed into the product of a unitary
    operation on $AC$ followed by a general operation on $BC$.
    \label{3autodecomp-fig}}
\end{figure}

A very similar argument can be applied to the three-system situation, in which
the global map $\cpmap{ABC}{}$ permits no information transfer from $B$ to $A$.
In this case, the map must have a unitary representation of the form shown in
Figure~\ref{3urepdecomp-fig}.
In general, the map $\cpmap{ABC}{}$ can {\em not} be written as the
composition of two maps acting on $AC$ and $BC$ alone, because these
subsystems each interact with the same environment $E$.  On the other
hand, if $\cpmap{A}{AC}$ is autonomous, then we can find a unitary
representation for it that does not include any interaction with the
external environment $E$.  This means that the global map $\cpmap{ABC}{}$
can be decomposed in this way
\begin{equation}
    \cpmap{ABC}{} = \left ( \unity^{A} \otimes {\cal F}^{BC} \right ) \circ
        \left ( \unity^{B} \otimes {\cal U}^{AC} \right )
\end{equation}
where ${\cal U}^{AC}$ is unitary.  This is shown schematically in
Figure~\ref{3autodecomp-fig}.

\section{Remarks}

Our discussion of dynamical locality in quantum mechanics has so
far been very general.

Classical cellular automata are idealized systems consisting of a
spatial grid of cells.  Each cell can only take on a finite number of
internal states at a series of discrete time steps.  During each time
step, the internal state of each cell is updated according to a rule
that is the same for every cell.  This rule takes as input the states
of the cell itself and a finite number of its immediate neighbors.
Thus, during a given time step, each cell only receives information
from the cells of its neighborhood.\cite{class-ca}

A quantum cellular automaton is a spatial grid of cells, each of which
is a quantum system described by a finite Hilbert space.  During each
discrete time step, the state of each cell is updated according to a
CP map which takes as input the joint state of the cell and its neighbors.
In other words, the update rule for a cell $A$ with neighbors $C$ is of
the form $\cpmap{A}{AC}$.  The system $B$ containing the rest of the grid
beyond the neighborhood does not influence the new $A$ state, so that
$B \noinfo A$.\cite{quant-ca}

There are complications in the quantum case that are not present in the
classical case.  For example, any local classical rule can be extended to
a global update rule for the entire grid.  However, this is not true for
a quantum cellular automaton.  It is possible to devise local CP maps
$\cpmap{A}{AC}$ that cannot be ``woven together'' in overlapping neighborhoods
to form a global CP map for the entire system.  An important (and to our
knowledge, open) question is what class of local CP maps can be linked
together consistently.

Armed with our analysis of locality, we can draw some interesting conclusions
about quantum cellular automata in general.  For instance, if the global
update map is unitary, then the local update maps $\cpmap{A}{AC}$ must
be autonomous.  This and other issues will be discussed in a later paper.

It is possible that our analysis could have application to the theory
of quantum cryptography.  The requirement that $B \noinfo A$ is a kind
of ``security condition'':  no information about secret system $B$ can
find its way to system $A$ (perhaps accessible to an eavesdropper), despite
the fact that both have interacted with $C$.  Our decomposition results
tell under what circumstances this condition holds exactly for arbitrary
initial states.

We also remark that our decomposition results are most intuitively
represented as statements about the rearrangement of a quantum circuit.
A complicated circuit can be replaced as shown in Figure~\ref{udecomp-fig}
if and only if the condition $B \noinfo A$ holds.  This may be a useful
idea for the design of quantum algorithms.

\section{Acknowledgements}

Both authors thank the Institute for Quantum Information at Caltech
for its hospitality, and one of us (Schumacher) gratefully acknowledges
the support of a Moore Distinguished Scholarship there in 2002--03.
Our thinking about this subject has benefitted decisively from many
conversations with Robin Blume-Kohout, Matthew Buckley, Michael Nielsen,
John Preskill and Reinhard Werner.

\end{document}